# Folding@Home and Genome@Home:

# Using distributed computing to tackle previously intractable problems in computational biology

Stefan M. Larson[1,2], Christopher D. Snow[1,2], Michael Shirts[1], and Vijay S. Pande[1,2,*]

[1]*Chemistry Department and* [2]*Biophysics Program, Stanford University, Stanford, CA 94305-5080*

[*]Corresponding author:

Vijay S. Pande

pande@stanford.edu

**Abstract**

For decades, researchers have been applying computer simulation to address problems in biology. However, many of these "grand challenges" in computational biology, such as simulating how proteins fold, remained unsolved due to their great complexity. Indeed, even to simulate the *fastest* folding protein would require decades on the fastest modern CPUs. Here, we review novel methods to fundamentally speed such previously intractable problems using a new computational paradigm: distributed computing. By efficiently harnessing tens of thousands of computers throughout the world, we have been able to break previous computational barriers. However, distributed computing brings new challenges, such as how to *efficiently* divide a complex calculation of many PCs that are connected by relatively slow networking. Moreover, even if the challenge of accurately reproducing reality can be conquered, a new challenge emerges: how can we take the results of these simulations (typically tens to hundreds of gigabytes of raw data) and gain some insight into the questions at hand. This challenge of the *analysis* of the sea of data resulting from large-scale simulation will likely remain for decades to come.

**Introduction**

The time is right for distributed computational biology. In the last few years, the influx of raw scientific data generated by molecular biology, structural biology, and genomics has completely outpaced the analytical capabilities of modern computers. Novel methods, algorithms and computational resources are needed to effectively process this wealth of raw information to continue the rapid rate of progress in modern biomedical science. For example, biomedical applications suggest the need to compute the structure, thermodynamics, dynamics and folding of protein molecules, the binding ability of drugs, and the key steps in biochemical pathways.

A biophysical approach to computational biology faces two major barriers: limitations in models (the level of abstraction used in representing the molecule, and the forces describing interactions) and limitations in sampling (how long a simulation can be run or how many different configurations can be visited; Cohen, 1996). The fundamental limitations in sampling can be directly attributed to the limits of current computers, which are typically 1000 to 100,000 times too slow for the demands of modern computational biology.

Recently, a new computing paradigm has emerged: a worldwide distributed computing environment, consisting of hundreds of thousands of heterogeneous processors, volunteered by private citizens across the globe (Butler, 2000). For example, SETI@home has accumulated over 400,000 years of single-processor CPU time in about three years. Distributed.net has used the power of this huge computational resource to crack DES-56 cryptography codes. In addition to the great scientific possibilities suggested by such enormous computing resources, the involvement of hundreds of thousands of non-scientists in research opens the door to new avenues for science education and outreach, in which members of the public become active participants.

Worldwide distributed computing holds out the promise of vast amounts of untapped computing power; the challenge is to correctly pose biological problems so as to be able to take advantage of these resources. Just as having 1000 assistants does not necessarily mean that work will be done 1000 times faster, it is often very difficult to subdivide large computational tasks into many smaller independent jobs. Many calculations, such as simulating protein folding kinetics with molecular dynamics, may seem ill suited for distributed computing, as parallel implications do not scale to more than tens to hundreds of processors with high speed networking, and are often unable to run on the heterogeneous processors and low speed networking found in distributed computing clusters. Thus, the great challenge for any distributed computing project lies in the development of novel algorithms, often requiring new ways to look at old problems in order to facilitate their calculation using distributed computing.

For example, dynamics of detailed atomic models of biomolecules are traditionally limited to the nanosecond timescale. Duan and Kollman have demonstrated that traditional parallel molecular dynamics (MD) simulations (essentially numerical integration of Newton's equations) can break the microsecond barrier (Duan and Kollman, 1998), provided one uses many tightly connected processors running on an expensive super-computer for many months. This type of calculation is poorly suited for worldwide distributed computing, where inter-processor communication is easily thousands of times slower, in both bandwidth and latency, than in a super-computer architecture. Clearly, novel algorithms must be developed for distributed computing. Our group has presented such an algorithm for atomistic biomolecular dynamics on distributed computing that reach orders of magnitude longer time scales than have previously been achieved (Pande, *et al*, 2002). At the heart of this method is the use of loosely connected multiple simulations to reach a greater sampling of conformations; analogous methods will likely play major roles in other computational biophysics problems, such as drug design. These methods, part of the Folding@Home distributed computing project, are discussed in further detail below. Of course, sampling is not the only limiting issue: the quality of models, such as the force-field potential energy parameters in molecular dynamics, is also critical. Fortunately, advances in sampling lead naturally to improvements in force-fields. By effectively solving the sampling problem, one can better understand the limitations of current models, and test new models with much greater statistical accuracy.

Earlier attempts to quantify biology on a molecular level have been hampered by the vast amount of computer power necessary to model the essential complexities. The explosion of genomic and proteomic data, combined with advances in computational algorithms and ever growing computational power, has opened the door to exciting biomedical advances. The combination of novel distributed computing algorithms with worldwide distributed computing will similarly serve to elevate computational biology to fundamentally new predictive levels.

**Challenges in computational biology**

The two dominant methods used in computational biology are physical simulation and informatic-based calculations. The appeal of physical methods is the dream of connecting our understanding of molecules and molecular interactions with sufficient precision and accuracy to describe complex biological phenomena. This approach poses three major challenges: models, sampling, and analysis (Figure 1).

Ideally, one would use models that are detailed and accurate enough to examine the desired question of interest, but not necessarily more so. When Newton wrote equations describing the motions of the planets, an atomistic model was neither necessary nor desired. For proteins, everything from lattice models, to off-lattice simplified alpha-carbon models, to fully atomistic models have been employed (for example, see Pande, 1998 and references therein). To test and examine protein design, lattice models have played a large role and have led to the development of the statistical mechanics theory of protein design (for example, see Pande, 2000 and references therein). On the other hand, for comparison of simulations of protein folding kinetics to experiment, it is likely that the protein-specific aspects of secondary structure driven by hydrogen bonding and side chain packing, aspects that are missing from lattice models, are critical for an accurate representation.

Sufficient sampling is likely the greatest source of inaccuracy in most areas of physically-based computational biology. In the case of protein folding kinetics, the problem of sampling manifests itself in terms of the limitations of simulating sufficiently long timescales: while proteins fold on the microsecond to millisecond timescale, typical simulations are limited to nanoseconds. Analogously, errors in free energy calculations are often dominated by entropic errors: molecular motions of interest are on the microsecond timescale or longer, whereas free energy calculations typically sample the pico- to nanosecond regime.

Clearly, there is an interplay between the challenges of modeling and sampling. More accurate models are typically more complex, and therefore sufficient sampling becomes more computationally demanding. This is where distributed computing can make the most significant contributions: given large computational resources, combined with algorithms of sufficient scalability, distributed computing can allow for both accurate models *and* sufficient sampling. The two case studies detailed below, Folding@Home and Genome@home, are good examples of this marriage of detailed models and thorough sampling.

Finally, the challenge of data analysis is perhaps less appreciated, especially in physical models. Simply reproducing reality (i.e. producing results with accurate models and sufficient sampling) has often been the overriding difficulty. However, distributed computing has the potential of breaking the barrier of sufficient sampling with accurate models. This leads to the side effect of producing an enormous set of data (easily hundreds of gigabytes) and shifts the focus from accurately reproducing reality to gaining new insight into the physical or biological processes of interest. Some approaches for dealing with and analyzing enormous datasets will be discussed below.

**The nuts and bolts of distributed computing**

Conceptually, client-server based worldwide distributed computing is straightforward. In practice, endless technical matters loom over every distributed computing project. In this section, we will detail the conceptual aspects and some of the pertinent details involved in running a distributed computing project.

Every distributed computing project begins with people across the world downloading and installing a piece of client software. The motivation for participation varies, from a lay interest in the project, to the desire to not waste idle computer cycles, to an amateur (or professional) interest in computers or the science involved. The client must contain everything that is necessary for the project to operate on the user's computer: networking, security, feedback, and the scientific code that is the heart of the project. Once installed, the client software downloads data (the "work unit") from the project's server, performs some computation, returns the results of the computation, and then restarts the process.

In terms of networking, there is a vast array of client-side issues that one must deal with, primarily due to the variety of security measures used to prevent unwanted internet connections. Common countermeasures include firewalls (which prevent certain types of incoming or outgoing networking communication) and proxies (which serve as intermediaries between the client computer and the outside world). Fortunately, due to the prevalence of web browsing, virtually all firewalls and proxies allow HTTP protocol transfers, over the typical TCP/IP ports used for HTTP (ports 80 and 8080). For this reason, most distributed computing client-server communication travels over the HTTP protocol, mimicking a webclient-webserver communication.

Security is critical in any distributed computing project. The validity of the final results rests on the integrity of the data returned by the clients. To ensure the integrity of this data, there are several standard computer science methods that can be applied. Digital signatures are often used to "sign" the data, ensuring that it was actually produced by the client software and was not altered before it was returned to the server. Digital signatures serve as a type of checksum, keyed by some long (e.g. 256 bit) string of data. If there was some change to the data, the digital signature check will fail and the server can determine that the data integrity was compromised. Alternatively, other cryptographic methods can be used, such as encryption. For most projects, there is no need for outright *secrecy*, just the guarantee of data *integrity*. Thus, checksums such as digital signatures are sufficient. It is important to note that while data integrity can be fairly easily maintained, secrecy is considerably harder.

User feedback is vital to any distributed computing project. Even if the technical aspects of the software work flawlessly, users are typically *donating* their computer time to the project. Often, their main motivation comes from some feedback which shows that their computers are actually performing the desired computation. This feedback can take many forms. Primarily, there is client-side feedback, such as graphics and information about the computation being performed, as well as server-side feedback, such as statistics web pages displaying how many work units a user has completed. Moreover, most projects allow users to pool their stats into "teams", and keep team rankings in addition to individual user rankings. Statistics play a very important role in distributed computing project, and serve as a primary means to help motivate volunteers to donate their computer time.

The most important aspect of the client is the computation itself. It is the heart of the calculation and the reason for a distributed computing infrastructure itself. An ideal calculation requires very little memory (capable of running on 32 MB machine) and little data transfer (~100K), but a large number of computer cycles (e.g. several hours to a few CPU-days per work unit). Of course, there are many calculations which do not meet these qualifications, and for which distributed computing may not be suitable. However, we stress that there are many calculations which on the surface appear to be unsuited to a distributed computing methodology, but with algorithmic advances, such as new server-side methods, can be rendered amenable to a distributed approach.

A distributed computing server faces its own challenges, especially for modern distributed computing applications, such as Folding@Home, which employ non-trivial server-side algorithms as an intrinsic part of its calculation (Shirts and Pande, 2001)  These challenges include the ability to handle a huge influx of data, the ability to handle heterogeneous clusters of client computers, fault tolerance for clients leaving the project, and the limits of low bandwidth/high latency communication.  An in-depth discussion of all the issues which go into running a large scientific distributed computing (e.g. server software design, hardware selection and maintenance, networking, and data archiving) is beyond the scope of this review. Let us instead focus on a specific server-side challenge from our work: the analysis of huge numbers of molecular dynamics trajectories.

Historically, the production of a handful of molecular dynamics trajectories has been the rule. Quantities of interest are computed through appropriate averaging over snapshots within a trajectory. Such calculations rely on the ergodic hypothesis, the assumption that the system time average will be equivalent to the ensemble average. The operative belief behind a molecular dynamics calculation is that the simulation has been run long enough to reach the ergodic limit. For biological systems, especially the folding of proteins, this approach is quite daunting. The generation of thousands of independent molecular dynamics trajectories should better sample the conformational space than a single trajectory of equivalent aggregate simulation time. However, the analysis of huge numbers of trajectories poses challenges of its own. Simple visual inspection becomes impractical when faced with thousands of independent MD trajectories. The storage of snapshots becomes unwieldy, and any analysis performed for each conformation becomes time consuming when repeated literally millions of times. It is important to determine what time resolution is necessary for the trajectory snap shots; we have used nanosecond snapshots for much of our work. When possible, however, it is preferable to analyze snapshots separated by 10-100 picoseconds. This ensures that the motion of the molecule is more continuous, with less information falling between the cracks. For instance, one might be interested in the order of hydrogen bond formation, or the timing of hydrophobic contacts among the sidechains.  Nanosecond snapshot resolution can obscure such detail.

Databases are a convenient utility for the rapid retrieval and cross analysis of the secondary structure and other computed qualities for each conformation. Even the storage of millions of protein conformations can be a challenge; most compression algorithms will not leave the data in a convenient form for further analysis. For a certain overhead, a database allows storage of the coordinates in a compact data structure at the desired precision. In principle, storage in a database will make for ready comparison between different projects. CGI interfaces to the database to perform analysis on the fly are fairly simple and database independent. Finally, we can look to the future when collaborative analysis of large ensembles of trajectories will be greatly aided by portable data structures that can be easily accessed online.

**Folding@Home: Simulating protein folding kinetics with distributed computing**

Protein folding (or misfolding) is a fundamental physical property of biomolecules. It plays a critical role in defining protein stability (including aggregation resistance), is the underlying cause of many diseases, and serves as a potential source of both technological innovations and limitations. The overarching research goal of the Folding@Home project (http://folding.stanford.edu) is a quantitative, predictive model of the folding process. Such a model would help elevate our understanding of protein folding such that current protein design methods (now limited to the design of particular native state structures) could be extended to include the design of stable, rapidly folding, aggregation-resistant proteins.

An appealing aspect of simulations is the ability to gain, in principle, an atomic level understanding of the mechanism by which proteins fold. Historically, however, simply running fully detailed, fully atomistic MD simulations was not a viable computational approach to the study of protein folding because even the fastest proteins fold on the seemingly intractable timescale of 10's of microseconds. Simulations are typically limited to the nanosecond timescale, a difference of 3 to 4 orders of magnitude. Thus, simulating protein folding in all-atom detail is computationally overwhelming for single processor approaches: at just ~1 ns per CPU-day, it would require decades on the fastest modern CPUs to simulate a few µs of reality. Consistent with this, the previous "record-holding" simulation (a 1 µs simulation of the villin headpiece) required months of CPU time on one of the fastest available supercomputers (Duan and Kollman, 1998).

Even with such great computational resources, fully detailed, atomistic MD simulations of folding appear unlikely.

However, we have recently developed and implemented a means of circumventing this computational bottleneck by employing distributed computing. Specifically, we have developed an algorithm that efficiently utilizes even the most highly asymmetric and distributed computer networks. We have used this algorithm ("ensemble dynamics") to perform atomistic biomolecular simulations that achieve orders of magnitude longer time scales than have previously been reached. In October 2000, we launched a distributed computing project dubbed Folding@home (Shirts and Pande, 2000; Shirts and Pande, 2001; Zagrovic, *et al*, 2001; Pande, *et al*, 2002). The Folding@home project incorporates molecular dynamics simulations and the appropriate client-server networking into a screen saver and makes this screen saver available to the public to download and run. Since that time, more than 40,000 participants have actively contributed to our simulations, accumulating 10,000 CPU-years in approximately 12 months. Our experience running Folding@home has demonstrated not only that people are willing and interested in participating in world-wide distributed efforts in computational biology, but that such an effort can provide significant advantages for the study of protein folding.

*The ensemble dynamics method*

Consider a system whose dynamics involves crossing free energy barriers. The dynamics of this system would be comprised of fluctuations in the initial free energy basin (e.g. the unfolded state), "waiting" for a fluctuation to push the system over the free energy barrier. How could we speed up the simulation of such a system? Voter applied such a mechanism for systems with energy barriers in materials science (Voter, 1998). We have extended this approach for application to systems with free energy barriers, such as protein folding.

In the case of a system with a single free energy barrier, the crossing time distribution (folding time distribution) is exponential:

$$P_1(t) = k \exp(-k\,t)$$

where $k$ is the folding rate. Traditionally this is taken to mean that one must simulate $1/k$ of real time in order to achieve a reasonable likelihood of capturing a folding event. What if, instead of simulating a single protein molecule, we simulated the folding of an ensemble of $M$ molecules in many parallel simulations and waited for the *first* simulation to cross the free energy barrier? The folding time distribution for this case (for the first one out of the simulations to fold in time $t$) is

$$P_M(t) = k \exp(-k\,t)\, M\, [1 - \int_0^t dt'\, k \exp(-k\,t')]^{M-1}$$

Where we multiply the probability that a particular simulation has folded ($k \exp(-k\,t)$) times the factor representing the possibility that any one of the simulations could be the first to fold ($M$) times the probability that the other $M-1$ simulations have not folded by time t. Simplifying the above, we get

$$P_M(t) = M\, k \exp(-M\, k\, t)$$

Thus, aside from some potential limitations described below, the folding rate for one of the $M$ simulations in the ensemble to fold is $M$ times faster than the folding rate for a single simulation. What would normally take 30 years on a single CPU could be simulated in 10 days using 1000 CPUs.

For a system with multiple barriers, we can still achieve a linear speed up in the number of processors, $M$ if we modify our method (Shirts and Pande, 2001). We cannot use the method above, since the system will linger in the intermediate states and the fastest possible folding time must include the time spent lingering in the intermediate states. Fortunately, we can reduce this lingering time via a slight modification of the ensemble dynamics method. When the first simulation undergoes a transition (for example, to an intermediate state), we restart all of the simulations from that intermediate state. This essentially reduces the problem to a chain of single barrier crossing problems, each characterized by a linear speed-up.

We have developed a novel method to identify these transitions. Since they are essentially crossings of free energy barriers, there should be a great variance in the energy during this event (analogous to a release of latent heat or a heat capacity peak in a first order phase transition). Indeed, energy variance peaks corresponding to transitions to meta-stable states have been seen in many simplified (Pande and Rokhsar, 1999a) and all-atom models of polymer folding (Pande and Rokhsar, 1999b). Operationally, we declare a transition if the energy variance exceeds some threshold value. While this value must be chosen empirically, we find the signal-to-noise of these peaks is sufficiently large that our results are robust to the precise definition of these parameters (see Shirts and Pande, 2001 for more details).

*Alternatives to ensemble dynamics: uncoupled trajectories.*

If the second most rate-limiting step is not too onerous computationally (e.g. is on the 1 to 10 ns timescale), one can take a much simpler approach to kinetics. Instead of coupling the simulations as in the ensemble dynamics method described above, one can simply run *M* independent trajectories. The fraction which should fold in time *t* is $F(t) = 1 - \exp(-kt)$ which can be approximated to $F(t) \sim k t$ for short *t*. For a $1/k = 10$ µs, we would expect that 1/1000 simulations of length $t = 10$ ns should fold. If we simulate 10,000 such trajectories, we should observe 10 folding events. This alternate method has the advantages of ease of analysis and interpretation, but it can get stuck in steps that are not rate-limiting experimentally, but are too slow computationally (*i.e.* ~10-100 ns). Thus, like the coupled trajectory case, there will be a minimum time that each simulation must complete in order to be able to cross the relevant free energy barrier, even on activated trajectories. This minimum time will greatly impact the scalability of the algorithm and is discussed below.

*Implications of these methods for simulating kinetics.*

Because of its extraordinary parallelization, ensemble dynamics puts the detailed, atomistic simulation of protein folding within reach. The ensemble dynamics method allows us to reach unprecedented simulation timescales even with just a relatively small cluster of processors. Using Tinker (with implicit solvent; Pappu, *et al*, 1998) or Encad (a fast MD algorithm for explicit solvent; Michael Levitt, personal communication) we can simulate 1 ns/day for a small, solvated polymer on a single Pentium III processor. Using even a 100-processor cluster, one can achieve 1 µs of simulation per week. The folding of the rapidly folding polymers described here (time-constants of ~5 µs) could thus be examined in approximately 10 weeks of CPU time. While this is computationally demanding, it is nevertheless quite feasible, putting the direct simulation of the folding of the smallest, most rapidly folding proteins within reach even of a somewhat large, but not uncommon computational resource.

While our coupled simulation method (ensemble dynamics) speeds transitions from the unfolded state to more native-like states, we also speed transitions from intermediate states back to the unfolded state. Similarly, ensemble dynamics cannot avoid "trapped" states, but it does escape them more quickly than traditional MD. Moreover, these traps and intermediates are an integral part of the dynamics we are studying. Thus, an ability to sample them *without* unproductively lingering in them provides a means of speeding up dynamics without significantly detracting from realistic trajectories.

Because the probability of a given simulation transiting the barrier in ensemble dynamics mirrors the single processor folding dynamics (except with a modified rate $k \rightarrow Mk$), ensemble dynamics provides a means of predicting folding rates. These predicted rates provide a clear and direct means of comparing our simulation results with an experimentally observed parameter. Our method also allows for the natural identification of folding intermediates (see below for a discussion of the viability of this method). As the transitions to intermediate states are demarcated by energy variance peaks, we can simply look back at our trajectories to identify at what times the simulation enters and exists these states. This should provide an ideal framework for Master Equation approaches (Ozkan*, et al*, 2001) to characterizing the rate-limiting steps in folding.

A significant advantage of ensemble dynamics is that it provides a means of identifying significant transitions without resorting to a potentially erroneous choice of reaction coordinate. Since we can identify transitions without looking at structural aspects of the polymers, we do not need to make any assumptions about the nature of the underlying dynamics. On the other hand, we can also use this method to test proposed reaction coordinates by using that degree of freedom to identify transitions.

It is important to note the potential limitations of the ensemble dynamics approach and their implications for protein folding research. While the method is considerably more tractable than traditional single processor or parallel processor (Duan and Kollman, 1998) MD approaches, it remains computationally demanding. There is an upper limit on the number of processors $M$ that can be used in a single simulation. Consider a protein folding with a 10 μs time-constant (*i.e.* $10^{10}$ MD iterations at 1fs per time step). If we employ our algorithm using $M = 10^{10}$ processors, we obviously will *not* fold a protein in a single MD iteration. There is an inherent limit to the scalability: no single MD simulation can fold a protein faster than the time it takes to transit the folding barrier. The natural resulting question is, "what is this fastest folding time?" The central hypothesis in our ensemble dynamics technique is that this fastest folding time (the time to cross the barrier) is considerably less (e.g. nanoseconds) than the overall folding time (microseconds). Our ability to fold a β-hairpin in the μs regime (see below) indicates that this is not a significant issue. Moreover, by examining the limitations of scalability (at which point adding processors no longer effectively speeds up our calculation), we can make a quantitative calculation of this fastest folding time: for example, if our simulations of a protein which folds in 10 μs scale up to 10,000 processors, then the fastest time is ~ 1 ns.

It is possible that distinct energy variance peaks might not be universal. This could impact the scalability of the ensemble dynamics method, since the inability to identify transitions means that trajectories will necessarily linger in metastable states. Alternatively, we could also look towards structural features to identify transitions: a transition is said to occur if there is some major change in tertiary or secondary structure identified by the number of native contacts or RMS deviation. Clearly the identification of transitions is the most difficult scientific issue we must address. Finally, if one uses completely independent trajectories, the identification of trajectories is not an issue. This is perhaps one of the strongest reason to use independent trajectories: due to its simplicity, the meaning of the data is absolutely clear.

*Results*

The Folding@Home client software used to generate the results below was based upon the Tinker molecular dynamics code (Pappu, *et al*, 1998). We simulated folding and unfolding at 30 ºC and at pH 7. We used the OPLS (Jorgensen, *et al* 1998) parameter set and the GB/SA (Qiu, *et al*, 1997) implicit solvent model. Stochastic dynamics (Pappu, *et al*, 1998) were used to simulate the viscous drag of water (γ = 91/ps), and a 2 fs timestep was used in our integration using the RATTLE algorithm (Andersen, 1983) to maintain bond lengths. 16 Å cutoffs with 12 Å tapers were used. Transitions were identified by the time resolved heat capacity spike associated with crossing a free energy barrier. To monitor the heat capacity during the simulation, we calculate the energy variance, and use the thermodynamic relationship $C_v = (\langle E^2 \rangle - \langle E \rangle^2)/T$; note that since we are using an implicit solvent, our "energy" is often called an "internal free energy" (the total free energy except for conformational entropy). Each PC runs a 100 ps MD simulation (one "generation"), calculates the energy variance within this time period, and then returns this data to the Folding@Home server. If the energy variance exceeds a preset threshold value (we used 300 (kcal/mol)$^2$ based upon earlier helical folding simulation results), the server identifies this trajectory as having gone through a transition, and then resets all other processors to the newly reported coordinates. Hundreds of generations are run in order to sufficiently sample the relaxation time needed to see folding.

How generally applicable and accurate is this method? To address this question, we applied the method to study the folding of a non-biological helix, a 12-mer of poly-phenylacetylene (PPA). This is a homopolymer with a simple sidechain that folds into a helix; it can be considered a non-biological analog of poly-alanine. Experimental studies demonstrate that helix formation occurs with a time-constant on the nanosecond timescale (Williams, 1996; Thompson, *et al*, 1997), suggesting that it is a computationally tractable system for the direct comparison of our simulation approach with traditional, single processor MD and with experiment. We find that ensemble dynamics produces mean folding times and folding time distributions consistent with those derived via traditional MD simulations (Figure 2). Moreover, our predicted folding time-constant of 10 ns accurately reproduces the experimentally observed value.

Eaton and coworkers (Thompson, *et al*, 1997) have experimentally demonstrated that alpha-helices form with time-constants of roughly 150 ns, well within the reach of our ensemble dynamics method. We have performed several folding simulations (Figure 3) of a 20-alanine helix, starting from a range of initial conformations (e.g. stretched versus collapsed). We find that the predicted time-constant (50 ± 50 ns) is quite comparable with that obtained experimentally. This provides further support that ensemble dynamics (and the OPLS force field) provide a sufficiently accurate description of the folding of isolated helices.

In addition to accurately reproducing experimentally observed rates, our simulations capture many of the detailed events observed experimentally during the folding of isolated helices. We observe, for example, events corresponding to the experimentally measured rapid structure formation in the amino-terminal part of the helix (based on the observation of amino-terminal fluorophore quenching in 10 ns by Eaton and coworkers (Thompson, *et al*, 1997) and the faster of two folding rates observed by Dyer and coworkers (Williams, 1996).

While the rapid folding of isolated helices is amenable to computational approaches, the question remains: is the folding kinetics of an isolated helix representative of the folding of a real protein? Clearly the study of helix formation will provide information relevant to the mechanisms underlying protein folding, but an isolated helix lacks an overall tertiary structure and thus the topomer search is probably not a significant contributor to the barrier that limits its rate of formation. What is the smallest polypeptide chain that can be said to adopt tertiary (or, perhaps more accurately, super-secondary) structure? Blanco *et al.* recently reported that a 16 residue hairpin excised from the β1 domain of protein G, unlike most small peptides, forms a unique, if only marginally stable, native state in solution (Blanco, *et al*, 1998). Of course, whether or not a 16-residue hairpin is a "super-secondary" structural element or actually exhibits tertiary structure is a matter of definition. It has been postulated that this hairpin is stable in solution due to the formation of a small, partially solvent-excluded hydrophobic core comprised of 2-3 side chains, suggesting that it is at least somewhat protein-like. A critical issue for our understanding of folding is that, while small, the hairpin still faces the Levinthal paradox of populating too many unfolded states to randomly search for the unique, native state in a biologically relevant time frame.

We have simulated the folding of the β-hairpin using ensemble dynamics on a 100-processor cluster, folding it in more than 20 independent simulations. Our predicted folding time ($4.7 \pm 2$ μs) is in close agreement with the experimental value (6 μs; Munoz, *et al*, 1997). Moreover, all trajectories folded by first forming the correct overall topology (possibly driven by hydrophobic interactions), then "zipping up" the native hydrogen bonds. This proposed mechanism is in agreement with the results of previous simulations which involved severe approximations, such as high temperature unfolding (Pande and Rokhsar, 1999b); for more details, see Figure 4 and (Zagrovic, *et al*, 2001).

Finally, while the experimental study of proteins has recently expanded from ensembles of proteins in solution to single molecule work, a shift in analysis that distributed computing allows is exactly the opposite. Rather than studying the formation and fluctuation of the alpha helical content of one trajectory we can consider the growth of the alpha helical content in an entire ensemble of proteins. We can imagine a new generation of ensemble calculations designed to predict observables such as the average radius of gyration, predicted NOEs or tryptophan fluorescence. Furthermore, distributed computing can increase our ability to characterize diffuse ensembles such as the unfolded state. Calculations that treat proteins as rigid objects, such as detailed electrostatics calculations to determine the interaction energy of a protein with solvent rely on the conformation of flexible surface residues. Here too, a thoroughly sampled system could increase our ability to perform accurate calculations.

**Genome@Home: Generating and using large libraries of designed protein sequences**

Protein folding involves finding the native three-dimensional structure for a particular amino acid chain in aqueous solution. The "inverse folding problem" (Pabo, 1983) seeks to define the set of sequences which can specifically form a stable protein with some target structure (i.e. the protein's "sequence space"). Traditional computational protein design algorithms seek to find amino acid sequences which are compatible with specific three-dimensional protein backbone structures. Applied on a large scale, computational protein design can provide important clues towards a solution of the inverse protein folding problem by broadly sampling the sequence space of known protein structures (Pande, *et al,* 2000). A clear understanding of sequence space and the nature of the protein sequence-structure relationship is critical for a wide range of fields in biological, medical, and chemical engineering research, such as protein engineering, protein evolution, and drug design.

Protein design has typically been used to predict one optimal amino acid sequence for an existing protein structure. This prediction can then be tested by synthesizing the polypeptide chain and experimentally assaying its stability, structure, and/or function (see for Pokala and Handel, 2001 for review). More recently, protein design algorithms have been used to design several sequences for a given structure. These sets of sequences can then be analyzed to make inferences about the nature of sequence space. Most of these studies have reached the conclusion that designed sequences tend to resemble the native sequence of the target protein structure (Koehl and Levitt, 1999; Kuhlman and Baker, 2000; Raha, *et al*, 2000).

Two major challenges lie in the way of fully developing protein design algorithms as mature tools for exploring sequence space: generation of sequence diversity and modeling of backbone flexibility. Meaningful sampling of sequence space requires running protein design algorithms multiple times, converging on minima from multiple starting points in the free energy landscape of sequence space. The second major hurdle in these efforts lies in modeling the inherent flexibility of the peptide backbone in the folded state. Since it is well known that natural proteins use small backbone adjustments to accommodate disruptive mutations (Eriksson, *et al*, 1992; Baldwin, *et al*, 1993), it is highly desirable to model this behaviour when computationally designing amino acid sequences. Incorporating backbone flexibility into computational protein design is a critical prerequisite for *de novo* protein design, where the exact structure of the resulting protein is not known. Unfortunately, the inclusion of backbone flexibility in the design process is extremely computationally demanding. Previous attempts at modeling backbone flexibility in design algorithms have been limited to coarse-grained variation of backbone structure parameters (e.g. relative arrangement of secondary or supersecondary structure element; Su and Mayo, 1997; Harbury, *et al*, 1998), or designing only a subset of residues in the target protein; Desjarlais and Handel, 1999). In all cases, only a small number of minimum-energy sequences for several proteins of interest were identified.

Though well-proven in single protein studies, the utility of protein design algorithms as tools for generating large libraries of designed sequences is hindered by their computational appetite. As well, studies incorporating backbone flexibility into the design process have been hindered by the increased computational complexity of annealing in conformation space on top of annealing in sequence space. To expand the scope of protein design algorithms for use in exploring sequence space, especially with the inclusion of backbone flexibility, requires much greater computational resources than previously available.

*Large-scale protein design using structural ensembles*

The accuracy of protein folding simulations depends on a combination of the models used and the extent of sampling of conformational space achieved. Similarly, protein design algorithms depend on using accurate models of the polypeptide backbone and the forces describing intramolecular and solvent interactions (generally quite similar to those used in MD simulations of protein folding), and extensively searching amino acid sequence space while keeping the conformation of the peptide backbone constant. In general terms, the chosen model provides a scoring function, and the design algorithm seeks to optimize an amino acid sequence to that function. Many of the standard minimization algorithms are used in protein design, with the main approaches being Monte Carlo, self-consistent mean-field approaches, dead end elimination, and genetic algorithms (Desjarlais and Clarke, 1998; Voigt, *et al*, 2000).

Given this algorithmic toolkit, the major obstacle to using protein design to learn about the sequence-structure relationship remains limited computational resources. Designing one optimized sequence for a single polypeptide backbone conformation using any of the methods described above takes on the order of a few CPU-hours (Voigt, *et al*, 2000). Inclusion of backbone flexibility (i.e. many backbone conformations) and extensive searching of sequence space increases computational demand by many orders of magnitude. Distributing computing can help bridge this gap.

The Genome@home project (http://gah.stanford.edu) uses an established design algorithm (Raha *et al*, 2000), deployed on a large scale through distributed computing. The server sends out a set of protein backbone coordinates and design parameters to the Genome@home client running on a user's computer. The client verifies the work unit, and runs the protein design algorithm. Upon completion of the sequence design, the results are verified by the client and sent back to the server, where the data is again verified, stored, and processed. Briefly, protein structures are created by modeling the placement of amino acid side-chain rotamers onto a fixed target backbone. The models are optimized by a sequence selection process that uses a genetic algorithm for finding an optimal sequence for the target structure. A diversity of sequences can be designed for the same target backbone, as the initial population of models is randomly assigned from a structure-dependent rotamer library, analogous to starting in a random point of sequence space. To mimic backbone flexibility, sequences are designed against an ensemble of 100 slight variants of each target structure, created through small perturbations of the dihedral angles of the target structure, within the global constraint of 1.0 Å root-mean-square deviation to the native structure.

Although designing to a structural ensemble is a fairly simple way of incorporating backbone flexibility, we have shown that it allows for a much broader search of sequence space than fixed-backbone methods. An initial study completed by Genome@home utilized over 3000 processors for 62 days to design a total of 187,342 sequences. The target set in this study consisted of all crystal structures in the Protein Data Bank of length less than 100 residues: 253 structures in total. Designing to a single, fixed backbone produces results very similar to other recently published studies addressing sequence analysis of designed sequences (Koehl and Levitt, 1999; Kuhlman and Baker, 2000; Raha, *et al*, 2000). Designing to a structural ensemble, however, produces a much greater diversity of sequences, and allows movement away from the region of sequence space immediately surrounding the native sequence. Homology searches against natural sequence databases show that the relevance of these sequences is not diminished (Figure 5). Initial tests showed that the increased diversity provided by Genome@home improves the utility of designed sequence libraries in fold recognition for structural and functional genomics.

**Future Perspectives**

We have introduced some of the basic concepts of the workings of distributed computing and have discussed the methodology employed in two particular case studies. Finally, below, we briefly speculate on the future of distributed computational biology. First and foremost, we predict that distributed computing will play an extremely important role in the future of scientific computing. Due to the remarkable price/performance ratio of Linux clusters on commonplace hardware, such clusters have become commonplace, and are perhaps the default computation platform in academia. However, such clusters were cutting edge computational methodology in the 1990's, and have recently become more common now that methods for building and utilizing such clusters are now well established. Similarly, we predict that distributed computing will itself cross over from bleeding edge to more common place methodology.

With the growing popularity of distributed computing as a research tool, are there sufficient resources for many separate research projects? There are hundreds of million personal computers on the Internet, and it is easily possible that 80-90% of this CPU power is wasted. If a distributed computing project involved 500,000 active users, as SETI@home currently reports, and half of all PC's now connected to the Internet participate, there would be capacity for hundreds of SETI-sized projects world-wide. The world's supply of CPU-hours is certainly not an inexhaustible resource, but it is very large, growing rapidly, and still virtually untapped.

Indeed, with thousands of PCs potentially available in university computer clusters, office desktops, and student dorms, virtually every university has supercomputer class resources waiting to be tapped. Analogous resources are also likely to be found in private companies or national labs. Since these computers are likely more homogenous than those found on the Internet as a whole and are likely connected with Ethernet networking, such resources have great potential. Indeed, due to this untapped power there are several initiatives which are also working to utilize this resource. "Grid computing" initiatives also work to combine multiple computers for use in large scale computation (Foster, 2002). Independent of the nature of the underlying software architecture used to connect processors and distribute computation, the same algorithmic issues remain: dividing a calculation among many processors connected by high-latency, low-bandwidth networking.

What is the future of distributed supercomputing calculations? In an ideal world, one would have access to a large computational resource (thousands to millions of processors) with high end networking (low latency, high bandwidth). IBM's Blue Gene project plans to provide just such a resource and to apply it towards protein folding. Blue Gene is proposed to have approximately 1 million processors connected by networking with microsecond latency and gigabyte bandwidth. This will allow for the scalability of traditional parallel molecular dynamics (where multiple processors are used to speed the calculation of a single MD trajectory) beyond the current state of the art of a few hundred processors (Duan and Kollman, 1998). However, scalability to a million processors may not be possible; in that case there would likely be

0.01 to 0.001 atoms per processor (compared with hundreds of atoms per processor in current parallel MD calculations). In this case, it is hard to devise algorithms to efficiently use these additional processors. One solution would be to take a hybrid approach by pushing traditional parallel MD to its limits (e.g. to 1,000 processors per trajectory) and then using the 1,000,000 processor computer to simultaneously calculate 1,000 trajectories (either in a coupled or uncoupled manner). Such hybrid approaches have the advantage of combining the efficiency of multiple parallelization methods in order to achieve unprecedented scalability. Unfortunately, such a resource will not be widely available. However, perhaps the true future of distributed computing lies in the combination of a large number of processors coupled by high-end networking. As broadband networking becomes more prominent, distributed computing may no longer be limited by networking communication, harkening a new class of problems and algorithms that could be tackled.

**Figure legends**

**Figure 1.** The primary challenges in computational biology: accurate models, sufficient sampling, and insightful analysis. Ideally, one would simply use the most accurate models available. However, there is typically a tradeoff between computational demands and the accuracy of ones models – by using accurate models, one typically limits the amount of sampling one can perform. Distributed computing allows one to have the best of both worlds: to employ large scale computational resources to allow sufficient sampling with highly accurate models. The case studies described here, Folding@Home and Genome@Home, both take this approach and use computationally expensive atomistic models, but still maintain sufficient sampling due to large scale parallel calculation. However, this brings up a third, perhaps lesser discussed challenge: how to gain insight from the sea of data which results. Indeed, large scale sampling with detailed models results in huge data sets. With the challenges of models and sampling met, we are now faced with this new challenge of winnowing down these data sets to their salient elements.

**Figure 2.** Shown is a comparison of two dynamics methods: probability distribution of folding at a given time for poly-phenylacetylene (PPA) via ensemble dynamics and traditional molecular dynamics simulations. As predicted mathematically, we see excellent agreement between the two distributions. Since PPA is computationally tractable, one can make this direct comparison to demonstrate the *quantitative* validity of ensemble dynamics. The observed time-constant, 10 ns, agrees well with experimentally observed rates (Yang, *et al*, 2000).

**Figure 3.** The ensemble dynamics method used to fold alpha helices. Shown above are trajectory data for simulations of the A-helix (left) and R-helix (right). Top: number of helical and beta-sheet units versus time (dotted line) and energy variance versus time. We see that peaks are associated with nucleation events. Bottom: Secondary structure versus time: red, yellow, and blue denote helices, beta sheets, and turns respectively. In both cases, we see nucleation events (corresponding to energy variance peaks). However, in the case of the R-helix, nucleation events did not occur at the arginine residues (R) and propagation typically was blocked at these residues (also seen in the other 7 runs we performed, data not shown).

**Figure 4.** The folding trajectory of an isolated β-hairpin. (a) A cartoon representation of the folding trajectory; the backbone of the peptide is represented as a gray trace and the core hydrophobic residues are shown in dot representation. (b) RMSD from the native structure of the hairpin, radius of gyration and the number of backbone-backbone hydrogen bonds formed. (c) The distance between key hydrogen bonding partners (green: trp43-val54, red: tyr45-phe52), and the minimum distance between trp43 and phe52 (black). Note that the distance between trp43 and phe52 reaches its native value before hydrogen bonding is established. d) Total potential, solvation and electrostatic energies ($E_{total}$, $E_{solvation}$ and $E_{charge}$) over the trajectory. The initial hydrophobic collapse of the unfolded peptide correlates with a sharp decrease in $E_{total}$. A significant deviation (around G170) of $E_{charge}$ and $E_{solvation}$ from their final values is correlated with the temporary breaking of the key tyr45-phe52 hydrogen bond. e) A summary of the key events along the folding trajectory (color code: yellow- high, violet- low). HB-ij denotes the distance between the hydrogen bonding partners *i* and *j*, min-kl denotes the minimum distance between residues k and l. Note that the establishment of the trp43-phe52 interaction is the earliest native-like structure formed.

**Figure 5.** Results of PSI-BLAST searches against the Protein Data Bank using sequence profiles generated from Genome@home designed sequences. 241 of the 253 structures (those that gave hits) are represented here, ranked along the x-axis by the E-value of the most significant hit obtained from that structure's designed sequence profile. Dark columns represent sequence profiles which gave hits against true homologues (true positives). Light columns identify sequence profiles which produced hits to non-homologues (false positives). A threshold of E < 1.0 gives an accuracy of 92% (176 of 186) for 74% (186 of 253) of all sequence profile searches. The black trace plots the average amino acid identity of each sequence profile to the native target sequence.

**Figure 1.**

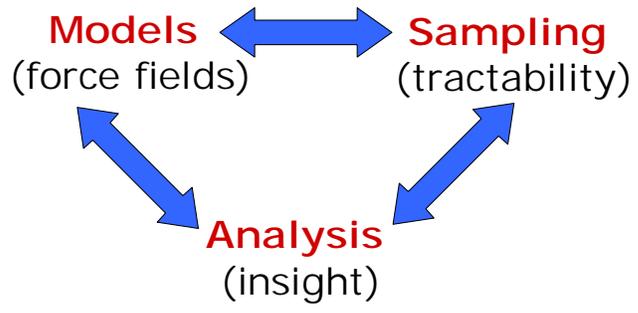

**Figure 2.**

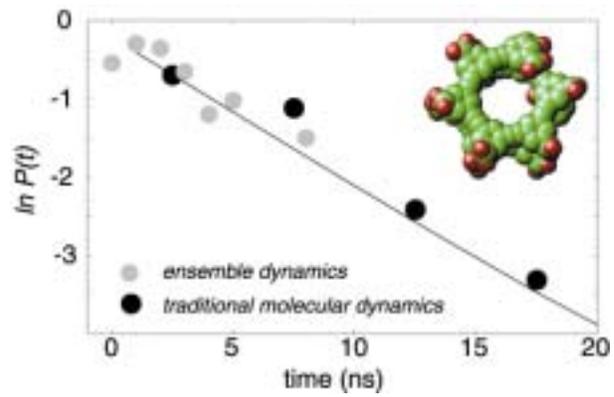

**Figure 3.**

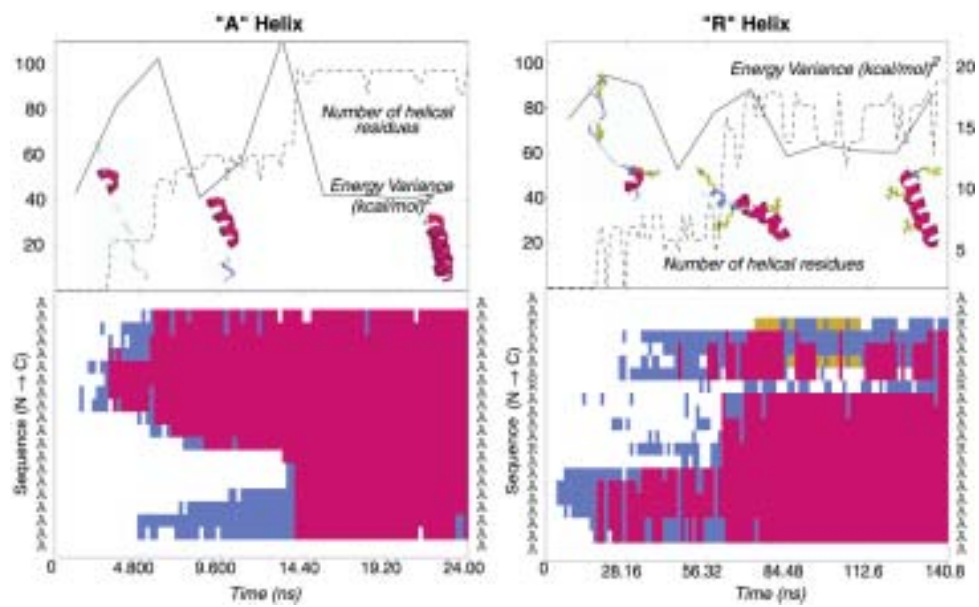

**Figure 4.**

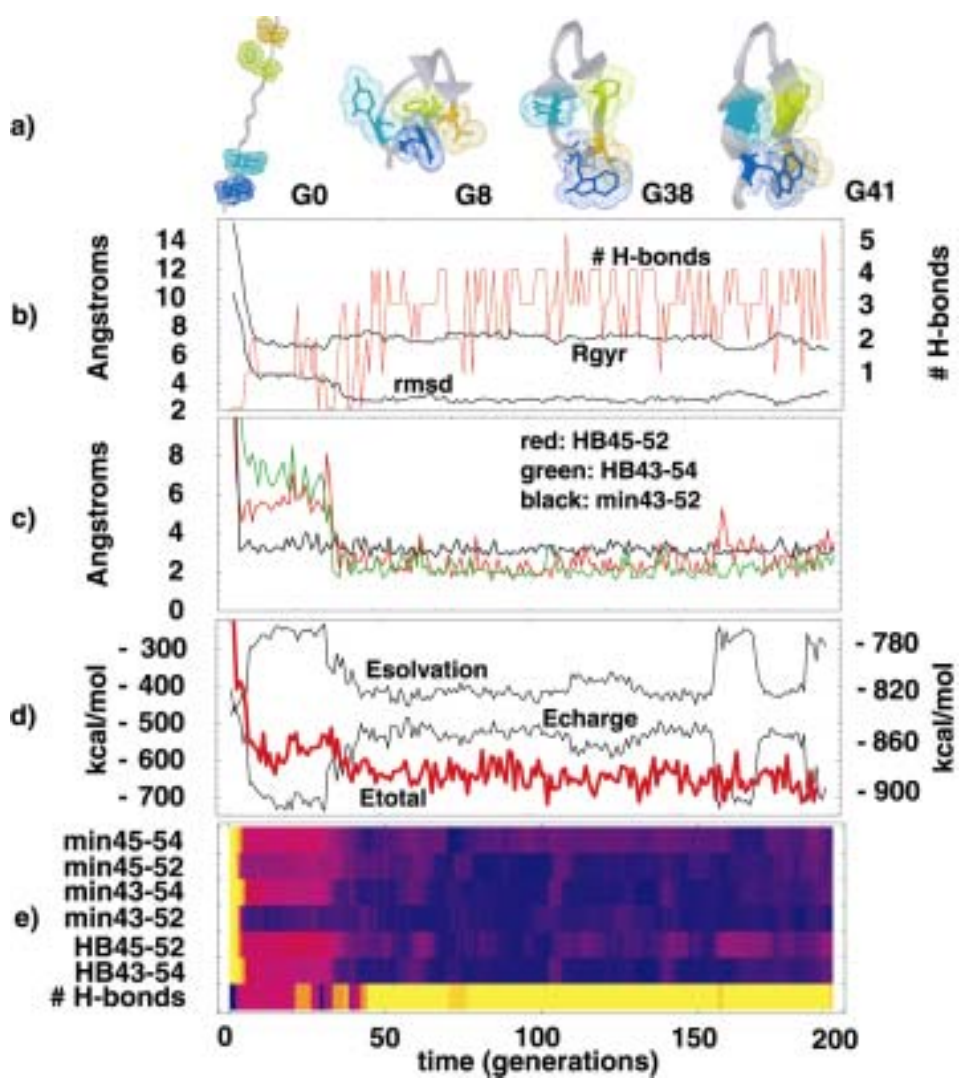

**Figure 5.**

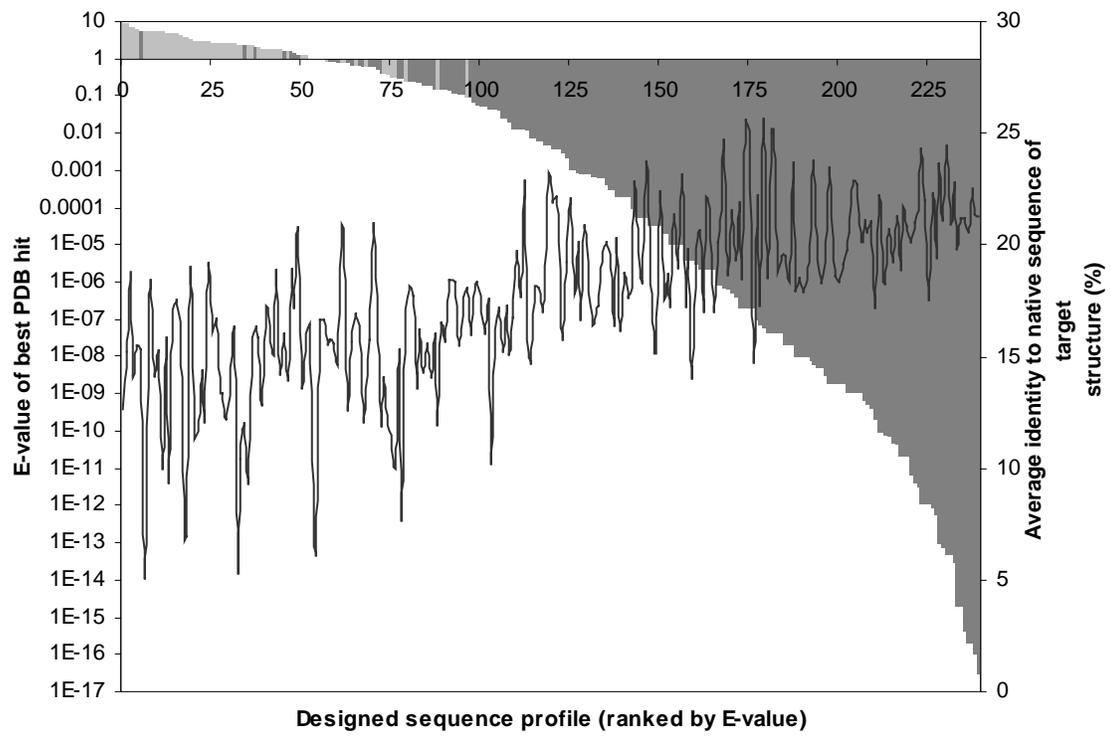